\documentclass[sn-mathphys,Numbered,pdflatex]{sn-jnl}% Math and Physical Sciences Reference Style

\usepackage{graphicx}%
\usepackage{multirow}%
\usepackage{amsmath,amssymb,amsfonts}%
\usepackage{amsthm}%
\usepackage{mathrsfs}%
\usepackage[title]{appendix}%
\usepackage{xcolor}%
\usepackage{textcomp}%
\usepackage{manyfoot}%
\usepackage{booktabs}%
\usepackage{algorithm}%
\usepackage{algorithmicx}%
\usepackage{algpseudocode}%
\usepackage{diagbox}
\usepackage{listings}%
\usepackage[normalem]{ulem}
\renewcommand{\vec}[1]{\mathbf{#1}}

\newcommand{\tens}[1]{\mathbf{\underline{#1}}}

\newcommand{\tb}[1]{_{\text{#1}}}
\newcommand*{\rhoC}[0]{\rho\tb{v1}}
\newcommand*{\rhoF}[0]{\rho\tb{v2}}
\newcommand*{\psiA}[0]{\psi\tb{d1}}

\newcommand*{\psiD}[0]{\psi\tb{d2}}

\newcommand*{\zetA}[0]{\zeta\tb{d1}}
\newcommand*{\zetB}[0]{\zeta\tb{b1}}

\newcommand*{\zetD}[0]{\zeta\tb{d2}}
\newcommand*{\zetE}[0]{\zeta\tb{b2}}

\newcommand*{\zetP}[0]{\zeta\tb{p}}
\newcommand*{\psiDi}[0]{\psi_{\text{d}i}}
\newcommand*{\hB}[0]{h_\text{b}}
\newcommand*{\hD}[0]{h_\text{d}}
\newcommand*{\hV}[0]{h_\text{v}}
\theoremstyle{thmstyleone}%
\theoremstyle{thmstyletwo}%

\theoremstyle{thmstylethree}%

\begin{document}

\title[Drops of volatile binary mixtures on brush-covered substrates]{Drops of volatile binary mixtures on brush-covered substrates}
\author*[1]{\fnm{Jan} \sur{Diekmann}}\email{jan.diekmann@uni-muenster.de}

\author[1,2]{\fnm{Uwe} \sur{Thiele}}\email{u.thiele@uni-muenster.de}

\affil[1]{\orgdiv{Institut f\"ur Theoretische Physik}, \orgname{Universit\"at M\"unster}, \orgaddress{\street{Wilhelm Klemm Str.\ 9}, \city{M\"unster}, \postcode{D-48149}, \country{Germany}}}

\affil[2]{\orgdiv{Center of Nonlinear Science (CeNoS)}, \orgname{Universit\"at M\"unster}, \orgaddress{\street{Corrensstr.\ 2}, \city{M\"unster}, \postcode{D-48149}, \country{Germany}}}

\abstract{
  \noindent We introduce a mesoscopic hydrodynamic model for drops of binary mixtures of volatile partially wetting liquids on brush-covered substrates, {i.e., we model the coupled dynamics of spreading, evaporation, imbibition, diffusion and partial demixing of the two volatile components across the three phases -- brush, drop and gas.}
  The formulation {of the model} as gradient dynamics on an underlying free energy functional allows us to systematically account for cross-couplings between the six scalar fields needed to describe the {dynamics of both components within each of the three phases.} The energy accounts for concentration- and brush state-dependent capillarity and wettability, miscibility of the two components within drop and brush, and entropy in the gas. Finally, the usage of the model is illustrated by employing numerical time simulations to study the {dynamics of a sessile drop.}
}

\keywords{adaptive wetting, polymer brush, mesoscopic hydrodynamics, volatile binary mixture, drop spreading}

\maketitle

\section{Introduction}\label{sec:introduction}
Overall, the statics and dynamics of the (de)wetting of simple nonvolatile liquids on rigid inert solid substrates are rather well understood \cite{BEIM2009rmp}. However, at present, flexible and adaptive substrates, e.g., (hydro)gels and polymer brushes, are {attracting much interest} \cite{LWBD2014jfm,BBSV2018l,AnSn2020arfm,HeST2021sm,HDGT2024l}. 
In general, films and drops of liquids on solid substrates are often modeled employing mesoscopic hydrodynamics in the form of thin-film (or lubrication) models that incorporate capillarity and wettability \cite{OrDB1997rmp,CrMa2009rmp,Thie2007chapter}. Particularly {instructive and} convenient is their formulation in the form of gradient dynamics on underlying energy functionals \cite{Mitl1993jcis,Thie2010jpcm,Thie2018csa,HeST2021sm,HDGT2024l}. Then, consistency between macroscopic and mesoscopic modeling {descriptions} is ensured by relations between macroscopic interface energies and {mesoscopic} wetting energies \cite{StVe2009jpm,TSTJ2018l,HDGT2024l}.

In the case of polymer brush-covered substrates \cite{LeMu2011jcp,YREU2018m,GaSo2019m,MeSB2019m,SmRB2020m,SHNF2021acis,RiCB2022aapm,SBNU2023l}, thin-film models in gradient dynamics formulation \cite{HDGT2024l} incorporate the free energy of the polymer brush, e.g., via an Alexander-de Gennes approach \cite{Alex1977jp,Genn1991crasi}. {In the simplest case, it accounts for mixing energy and entropy of the brush polymer and the imbibing liquid \cite{Somm2017m,ThHa2020epjt}. As the polymers are grafted to the supporting solid, their translational entropy does not contribute and is replaced by entropic stretching.} The resulting thin-film models may be employed to investigate the spreading of drops of nonvolatile \cite{ThHa2020epjt,HDGT2024l} and volatile \cite{KHHB2023jcp,HDGT2024l} liquids on polymer brushes, the stick-slip motion of three-phase contact lines advancing over an initially dry brush \cite{GrHT2023sm,HDGT2024l}, as well as sliding drops on such substrates \cite{HDGT2024l}. {Thereby, evaporation and vapor dynamics are incorporated into the gradient dynamics model following an approach recently proposed for drops of volatile simple one-component liquids in a gap \cite{HDJT2023jfm}.}

{Here, we provide a major extension of such models by incorporating the description of liquid mixtures, i.e., for a binary mixture both components are present in brush, drop and vapor. Although, this is of large present interest \cite{Somm2017m,YBUF2019m,SmRB2020m,SBNU2023l}, to our knowledge, the coupled dynamics of spreading, evaporation, imbibition, diffusion and (partial) demixing of the two volatile components across the three phases -- brush, drop and gas -- has not yet been modeled. We emphasize that the gradient dynamics-based approach ``semi-automatically'' guarantees thermodynamic consistency and nevertheless incorporates the relevant dissipation channels. Otherwise, this consistency might easily get lost when modeling such complex interface-dominated physico-chemical systems. Note, however, that we restrict our attention to isothermal settings as gradient dynamics-based thin-film models that incorporate heat transport are, to our knowledge, not yet available.}

{Note, however, binary mixtures of nonvolatile liquids on rigid inert solids are already described by mesoscopic hydrodynamics models \cite{NaTh2010n,Thie2011epjst,ThTL2013prl,XuTQ2015jpcm}. This allows, e.g.,} to investigate dewetting processes that are triggered by the coupling of fluctuations in film thickness and concentration \cite{ThTL2013prl}. Surface activity due to insoluble or soluble surfactants may also be incorporated \cite{ThAP2016prf, TSTJ2018l}.

{As mentioned above, evaporation and vapor diffusion for volatile liquids are accounted} for in thin-film modeling by confining the system into a small gap between two parallel plates as then vapor diffusion is predominantly lateral while concentrations equilibrate fast across the gap \cite{HDJT2023jfm}. {In contrast to earlier thin-film models including evaporation, see Refs.~\cite{OrDB1997rmp,CrMa2009rmp} and the discussion and references in the introduction of Ref.~\cite{HDJT2023jfm}, this allows one to describe the full parameter range from phase transition-limited to diffusion-limited evaporation/condensation. An application of the evaporation model of Ref.~\cite{HDJT2023jfm} to droplets of a simple weakly volatile liquid spreading on a brush \cite{KHHB2023jcp} shows that even a small lateral vapor leakage from a closed system} may result in the formation of a long-lived stationary halo of macroscopic extension in the brush profile.

The model is introduced in section~\ref{sec:model} by detailing the six scalar fields representing the amount of the two substances in the three phases brush, drop and gas, and the corresponding dynamic equations. {Details regarding the underlying energy functional and mobility functions are provided in Appendices~\ref{sec:appendixA} and~\ref{sec:appendixB}, respectively.} Using the established model, section~\ref{sec:results} presents exemplary results obtained by numerical time simulations. Finally, {section~\ref{sec:conclusion}} summarizes our results, points out possible applications of the developed model, and proposes future extensions that may alleviate present limitations. 
\section{Mesoscopic hydrodynamic model in gradient dynamics form}\label{sec:model}
\begin{figure}[htb]
	\centering
	\includegraphics[width=0.9\textwidth]{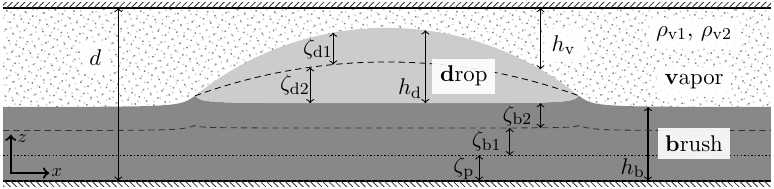}
	\caption{Sketch showing a drop of thickness $h_\mathrm{d}$ of a binary volatile liquid mixture on an adaptive substrate, namely, a polymer brush of thickness $h_\mathrm{b}$, confined in a narrow gap of width $d$. The amounts of the two components in each phase - brush, drop and gas - may be given as effective layer heights $\zeta_{\mathrm{d}1}, \zeta_{\mathrm{d}2}, \zeta_{\mathrm{b}1}, \zeta_{\mathrm{b}2}$ and vertically averaged number densities per volume $\rhoC, \rhoF$ or as particle number densities per substrate area $\psi_i$ (see main text). In the sketch the $\zeta$'s are illustrated as 'layering'.}
	\label{fig:full-system-sketch}
\end{figure}

We consider an isothermal system as sketched in Fig.~\ref{fig:full-system-sketch}: A drop of a binary mixture of volatile liquids is located on a swelling polymer brush. {Described are} the amounts of the two liquid components in the drop and brush by effective heights $\zeta_{\mathrm{d}i}$ and $\zeta_{\mathrm{b}i}$ (with $i=1,2$), respectively, the thickness of the drop is $\hD=\zetA+\zetD$ while the brush thickness is $\hB=\zetB+\zetE+\zetP$. Here, $\zetP$ is the uniform and constant effective height of polymers, i.e., the dry brush height $\zetP=\sigma N l\tb{K}$, with $\sigma$ the relative grafting density, $N$ the number of monomers per polymer chain and $l\tb{K}$ the Kuhn length. 
We consider the system to be positioned inside a narrow gap between two horizontal plates of distance $d$, therefore, the height of the gas phase is $\hV=d-\hD-\hB$. 

Thin-film models for liquid films and drops involving mixtures are conveniently written in gradient dynamics form employing (effective) heights \cite{ThTL2013prl,XuTQ2015jpcm,ThAP2016prf,TrJT2016ams}. 
However, when vapor is incorporated, it is preferable to express the amounts of each component in each phase - brush, drop and gas - as particle number densities per substrate area $\psi_{i}(\vec{r},t)$, where $\vec{r}=(x,y)^T$ denotes the substrate coordinates \cite{KHHB2023jcp,HDGT2024l}. Here, we use the {six fields} $(\psi_1,\psi_2,\psi_3,\psi_4,\psi_5,\psi_6)=(\rho_{\mathrm{l}1}\zetA, \rho_{\mathrm{l}1}\zetB, \rho_{\mathrm{v}1}\hV, \rho_{\mathrm{l}2}\zetD, \rho_{\mathrm{l}2}\zetE, \rho_{\mathrm{v}2}\hV)$ where the $\rho_{\mathrm{l}1}$ and $\rho_{\mathrm{l}2}$ denote reference number densities per volume of the pure liquids 1 and 2, respectively. {We emphasize that the system geometry is identical to the one presented in \cite{HDGT2024l,KHHB2023jcp}. However, here we provide a major extension of earlier work by incorporating the description of a binary mixture, i.e., two substances (instead of only one) contribute to the dynamics in each bulk phase.}

Employing the introduced fields $\psi_i$ in the gradient dynamics model yields the generic form \cite{ThAP2016prf,Thie2018csa,HDGT2024l}
\begin{align}\label{eq:gradDyn}
	\partial_t \psi_{i} =
	\nabla \cdot \left[\sum_{j=1}^{6}Q_{ij}\nabla\frac{\delta \mathcal{F}}{\delta \psi_{j}}\right]
	-\sum_{j=1}^{6}M_{ij}\frac{\delta \mathcal{F}}{\delta \psi_{j}}
	\qquad i=1,\dots,6
\end{align}
where $\tens{Q}$ and $\tens{M}$ are positive semi-definite and symmetric $6\times6$ matrices representing the mobilities for the conserved and nonconserved parts of the dynamics, respectively. {The individual components of both mobility matrices are discussed in Appendix~\ref{sec:appendixB}.} Note that the mobilities of Refs.~\cite{ThTL2013prl,XuTQ2015jpcm,KHHB2023jcp,HDGT2024l} are recovered as limiting cases. The functional $\mathcal{F}$ corresponds to the underlying free energy, i.e., the functional derivatives $\delta \mathcal{F}/\delta \psi_i$ represent chemical potentials. Their gradients within each phase and differences between phases drive the conserved (advection, diffusion) and nonconserved (evaporation/condensation, imbibition/desiccation) parts of the dynamics, respectively. We use
\begin{align}\label{eq:FullFreeEnergy}
	\mathcal{F}&=\int_{\Omega}\left[ 
	\underbrace{f_\mathrm{b}(\zeta_{\mathrm{b}1}, \zeta_{\mathrm{b}2})}_{\text{brush bulk}} + 
	\underbrace{f_\mathrm{d}(\zeta_{\mathrm{d}1}, \zeta_{\mathrm{d}2})}_{\text{drop bulk}} + 
	\underbrace{\hV f_\mathrm{g}(\rhoC,\rhoF)}_{\text{gas bulk}} +
	\underbrace{\xi_\mathrm{dg}\gamma_\mathrm{dg}(\zeta_{\mathrm{d}1}, \zeta_{\mathrm{d}2})}_{\text{drop-gas interface}} \right.
	\nonumber\\
	& + 
	\underbrace{\xi_\mathrm{bd}\gamma_\mathrm{bd}(\zeta_{\mathrm{b}1}, \zeta_{\mathrm{b}2},\zeta_{\mathrm{d}1}, \zeta_{\mathrm{d}2})}_{\text{brush-drop interface}} 
	\left.+\underbrace{\xi_\mathrm{bd}f_\mathrm{wet}(\zeta_{\mathrm{b}1}, \zeta_{\mathrm{b}2},\zeta_{\mathrm{d}1}, \zeta_{\mathrm{d}2})}_{\text{wetting}}
	\right] \mathrm{d}^2r
\end{align}
where $\xi_\mathrm{dg}=\sqrt{1+|\nabla(\hD+\hB)|^2}$ and $\xi_\mathrm{bd}=\sqrt{1+|\nabla\hB|^2}$ denote the metric factors for the two interfaces, and we use the $h$'s, $\zeta$'s and $\rho$'s as abbreviations for lengthy expressions only involving dependencies on the fields $\psi_i$. In total, there are bulk energies for the three phases, two interface energies and one wetting energy. The latter ultimately accounts for the brush-gas interface energy within the mesoscopic description \cite{TSTJ2018l, HDGT2024l}. The specific forms of the individual contributions extend cases already considered the literature while recovering them as limiting cases. Examples include drops of simple nonvolatile and volatile liquids on polymer brushes \cite{HDGT2024l}, and drops of nonvolatile mixtures on rigid substrates \cite{ThTL2013prl}.
{The individual expressions of all contributions to  $\mathcal{F}$ are discussed in Appendix~\ref{sec:appendixA}.} 

Alternatively, instead of the thermodynamic formulation as gradient dynamics \eqref{eq:gradDyn}, one may present the same model in hydrodynamic form in terms of advective and diffusive fluxes and transfer rates between the phases. As the formulation is fully equivalent we abstain from showing it here, {but see Ref.~\cite{HDGT2024l} for details regarding the ``translation'' between the two formulations (at the example of a simple liquid)}.
\section{Results}\label{sec:results}
Having completed the model, we next illustrate its usage by briefly discussing the relaxational spreading, evaporation and imbibition dynamics of a binary mixture of volatile liquids on an initially dry polymer brush. We employ numerical time simulations based on the fully adaptive (time and space discretization) finite-element methods provided in the package \textit{oomph-lib} \cite{HeHa2006} and consider the full-curvature formulation of our model as described in section~\ref{sec:model} and the appendices. We remind the reader that our gradient dynamics approach only resolves lateral concentration gradients within the mixture, i.e., densities are vertically averaged within brush, drop and gas, respectively. 

\begin{figure}[ht]
	\includegraphics[width=\textwidth]{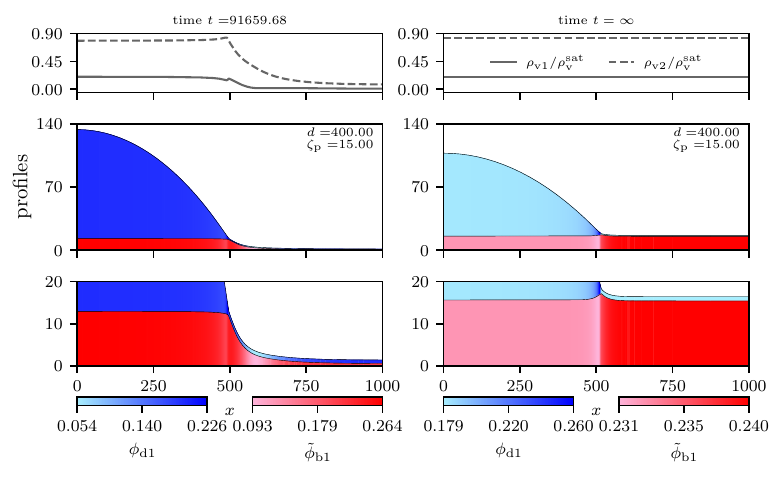} 
	\caption{Snapshots from a simulation of a spreading, evaporating and imbibing drop of a binary mixture of volatile partially wetting liquids on an initially dry polymer brush in 1d. On the left hand side an intermediate state is shown while the right hand side presents the final equilibrium state. Top, center and bottom {panels} give concentration/thickness profiles in gas, drop and brush, respectively. For initial state, parameters, and a discussion of further details and main features see the main text and appendix~\ref{sec:appendixC}. Lateral boundary conditions are applied and only half the computational domain is shown.
	\label{fig:simulation1}
	}
      \end{figure}

      Fig.~\ref{fig:simulation1} presents snapshots from an exemplary simulation with an initial drop of homogeneous concentration $\phi^0_\mathrm{d1}=0.2$ and contact angle of $\theta^0\approx 45\,^\circ$ on a nearly completely dry brush ($\phi^0_\mathrm{bp}>0.999$). The initial contact line position is $x_\mathrm{cl}^0\approx 365$.
      The equilibrium contact angle of such a mixture on the dry polymer would be $\theta_\mathrm{ref}\approx26.3\,^\circ<\theta^0$, i.e., when disallowing particle exchange between phases (all $M$'s equal zero) the drop will just spread. However, here we allow all transfers to occur and spreading will compete with evaporation and imbibition. The complete set of employed parameters is given in appendix~\ref{sec:appendixC}. The left and right hand column in Fig.~\ref{fig:simulation1} shows the state during the dynamic process at $t\approx 9.1\cdot10^{4}$ and after the system has converged to its equilibrium state, respectively. Thereby, the top panels give the state of the gas layer in terms of relative humidities for both vapors ($\rho_{\mathrm{v}i}/\rho^\mathrm{sat}_{\mathrm{v}}$), see Appendix~\ref{sec:appendixA} for the definition of $\rho^\mathrm{sat}_{\mathrm{v}}$, the center panels show the drop and brush height and concentration profiles, while the bottom panels magnify the thickness range of the brush. The liquid concentrations $\phi_\mathrm{d1}$ and $\tilde\phi_\mathrm{b1}$ within the drop and the brush, respectively, are encoded as respective blue and red shading (see corresponding color bars underneath at bottom, and note the different scales for left and right panels). In the brush, $\tilde{\phi}_\mathrm{b1}=\zetB/(\zetB+\zetD)$ is the concentration with respect to only the imbibing liquid mixture (not including the polymer). The volume fraction of polymers may be extracted from the presented brush height (displayed with a shift of~$\zetP$) using $\phi_\mathrm{bp}=\zetP/h_\mathrm{b}$. The actual part per volume concentration of the individual liquids within the brush can then be obtained as $\phi_{\mathrm{b}i}=\tilde{\phi}_{\mathrm{b}i}(1-\phi_\mathrm{bp})$.

      As the drop spreads on the brush-covered substrate, the two components of the mixture are absorbed into the brush and evaporate in a differentiated manner as controlled by the underlying energies. As a result, thickness profiles of drop and brush change as do the concentration profiles in drop, brush and gas. The snapshot on the left of Fig.~\ref{fig:simulation1} approximately captures the instant when spreading has ceased and the dynamics becomes dominated by evaporation and imbibition alone.\footnote{For spreading and evaporating drops of simple liquids on solid substrates this state has been analyzed in Ref.~\cite{ToTP2012jem}.} The vapor above the drop is nearly saturated and the brush has strongly swelled underneath the drop with a macroscopic halo in the brush profile outside the drop (cf.~\cite{KHHB2023jcp}). This halo is matched by corresponding vapor concentration gradients in the gas phase above the brush beyond the drop. Careful inspection shows that the vapor density of liquid 1 decays faster than the one of liquid 2 indicating that the concentration in the brush may change non-monotonically (assuming approximate local equilibrium between vapor and brush state). Indeed, the lower panel on the left indicates such a behavior with a minimum in $\tilde{\phi}_\mathrm{b1}$ at $x\approx575$.

      When equilibrium is reached (right panels of Fig.~\ref{fig:simulation1}), the vapor concentrations are homogeneous across the system (diffusion has ceased), and approximately fulfill Raoult’s law (partial pressures of components in the mixed vapor phase equal the vapor pressure above a pure liquid weighted by their mole fraction \cite{AtkinsPaula2010}) with respect to the mixing ratio of an ideal liquid mixture in a drop. A small deviation ($\sum_{i=1}^{2}\rho_{\mathrm{v}i}/\rho^\mathrm{sat}_{\mathrm{v}}\approx1.03>1$) results from our consideration of non-ideal mixtures ($\chi_i\neq0$).
Also the brush has reached its swelling equilibrium that only slightly differs underneath and beyond the drop in thickness (cf.~section ``Sorption isotherm of a polymer brush'' of \cite{HDGT2024l}) and concentration. Note that the transition between the two plateau values of the concentration seems to be nearly step-wise - this should in the future be amended by incorporating an energetic penalty for steep concentration gradients. The final equilibrium brush profile nicely shows that a wetting ridge has emerged in the three-phase contact line region. There, the concentration within the drop and brush are both locally influenced by wettability: coming from the drop side where $\phi_\mathrm{d1}=0.182$ [$\tilde\phi_\mathrm{b1}=0.119$] the concentration $\phi_\mathrm{d1} $ [$\tilde\phi_\mathrm{b1}$] first increases [decreases] in the contact line region before it decreases [increases]  again toward the adsorption layer [brush] outside the drop to $\phi_\mathrm{d1}=0.179$ [$\tilde\phi_\mathrm{b1}=0.121$]. Overall, one can see that concentration differences across the system are smaller in the final equilibrium state than in the course of the dynamics. 
Overall, the example simulation shows all features of the process expected on general grounds and allows one to study many specific properties of the interacting dynamical processes consistently driven by the underlying energy.
\section{Conclusion}\label{sec:conclusion}
{We have introduced a mesoscopic hydrodynamic model for drops of a binary mixture of volatile liquids on brush-covered substrates in a narrow-gap geometry. Our work corresponds to a major extension of earlier work on gradient dynamics-based thin-film descriptions of drops of volatile simple liquids in a gap \cite{HDJT2023jfm} as well as of nonvolatile and volatile simple liquids on brush-covered substrates \cite{HDGT2024l}. Although, here we have only presented the full-curvature version of the model (indicated by the usage of the exact metric factors), a long-wave version is easily obtained by approximating the metric factors for small interface slopes, see, e.g., pertinent discussions in Refs.~\cite{Thie2018csa, HDGT2024l}.}
  
We have shown that the formulation as a gradient dynamics on an underlying free energy functional allows one to 'quasi-automatically' account for all energetic cross-couplings between the individual fields and the resulting dynamic processes. In particular, the needed six scalar fields represent the two liquid components {of the mixture within the brush and the drop as well as the two components of the vapor in the gas phase.} All contributions to the energy have been detailed in Appendix~\ref{sec:appendixA} while the matrices of mobility functions for the conserved and nonconserved parts of the dynamics are discussed in Appendix~\ref{sec:appendixB}. The energy accounts for concentration- and brush state-dependent capillarity and wettability, miscibility of the two components within drop and brush, as well as entropy in the gas phase. The developed model may account for the coupled spreading, absorption, imbibition, diffusion, evaporation/condensation, demixing/mixing and swelling dynamics that occurs when a drop {of binary mixture} is placed on a dry polymer brush. We have briefly illustrated this by employing an exemplary numerical time simulation of such a process. It has shown that the dynamic model seems to well describe the intricate coupled processes and may in the future serve in detailed studies of a variety of particular processes and phenomena related to the interaction of volatile mixtures and brushes. Examples include the experimental study of the influence of saturated vapors of different liquids on the wetting properties of brush-covered substrates that show atmosphere-dependent swelling \cite{SBNU2023l}. Similarly, the presence of different organic vapors in the gas phase influences contact angle hysteresis for and resulting velocity of water drops on tilted brush-covered substrates as compared to the case without such vapors \cite{LHKS2022am}.

{Note, however, that the developed model only represents a first step in modeling the dynamics of droplets of mixtures of volatile liquids on adaptive substrates. Although, here we have shown that the gradient dynamics form of mesoscopic hydrodynamics is well suited to extend the reach of thin-film models toward liquid mixtures, a number of improvements of the presented basic model should be pursued in the future:}
(i) One may refine the brush energy accounting for more subtle interactions between brush, liquid mixture (and vapors). For instance, using the energy proposed in \cite{Somm2017m} should facilitate studies of the dynamics of various processes related to co-nonsolvency \cite{ENLA2010l,DuFD2016jpcb,YCKK2016p,SHNF2021acis,SBNU2023l}. {(ii) The present model does not account for  advective flows driven by solutal Marangoni forces resulting from concentration dependencies of interface energies. However, they are highly relevant for sessile drops of mixtures \cite{LLDL2022jfm} or the merging of drops of different liquids \cite{KaRi2014jfm}. They may be incorporated into the presented model by employing two additional fields that describe the enrichment of components at the brush-drop and the drop-gas interface, respectively.} This is similar to the incorporation of a surfactant concentration field as done in the gradient dynamics formulations in Refs.~\cite{ThAP2012pf,ThAP2016prf}. In principle, following the approach in section IV.B of Ref.~\cite{ThAP2016prf} one may then apply an approximation to again eliminate the additional fields still keeping the Marangoni effect (at the price of slightly breaking the gradient structure). (iii) Beside the bulk energies one may also refine the concentration-dependencies of interface and wetting energies beyond the here-employed linear interpolations between known limiting cases (see, e.g., \cite{ATTG2024pre}). There, a discussion is needed that considers relations between the concentration dependencies of interface and wetting energies and the bulk mixture energies.
(iv) Phase separation within the liquid mixture may result in micro-phase separation and very steep concentration gradients (as already visible in the contact line region in the studied example). To avoid such steep gradients one may incorporate an energetic penalty for concentration gradients, e.g., by using square-gradient terms in the energies for drop and brush phase similar to Cahn-Hilliard type energy densities \cite{ThTL2013prl}. (v) As in \cite{HDGT2024l}, we have assumed that transport of the two liquids within the brush is by diffusion only. This neglects any dynamic coupling between the lateral motion of the liquids within the brush and within the drop. Instead, one may model the motion within the brush similar to a (brush state-dependent) porous layer or, alternatively, introduce a brush state-dependent effective slip at the brush-drop interface (see, e.g., \cite{ThGV2009pf}). The former would add further nondiagonal terms to $\tens{Q}$ while the latter would only amend existing nonzero components of $\tens{Q}$. {(vi) Our work has assumed an isothermal setting, although effects driven by latent heat and subsequent diffusion of heat are often relevant when considering evaporation \cite{WiDA2023arfm}. To alleviate this limitation one would need to consistently incorporate the transport of heat into the gradient dynamics formalism. To our knowledge, this has not yet been done.}

Note, finally, that the proposed model does not only allow one to describe dynamic processes for binary drops on adaptive substrates formed by polymer brushes. It may also serve as a blueprint for the development of similar models for most other types of adaptive substrates discussed in \cite{BBSV2018l}.

\backmatter
\bmhead{Supplementary information}

\bmhead{Acknowledgments}
This work was supported by the Deutsche Forschungsgemeinschaft (DFG) within SPP~2171 by Grant No.\ TH781/12-2. We acknowledge fruitful discussions on adaptive viscous and viscoelastic substrates with Christopher Henkel and Jacco Snoeijer, on evaporation modeling with Simon Hartmann as well as on the interaction of liquids and polymer brushes with Daniel Greve and the group of Sissi de Beer at Twente University.

This version of the article has been accepted for publication, after peer review but is not the Version of Record and does not reflect post-acceptance improvements, or any corrections. The Version of Record is available online at: \url{http://dx.doi.org/10.1140/epjs/s11734-024-01169-4}.
\section*{Declarations}
\textbf{Data Availability Statement}
\\
The underlying data needed to reproduce shown results are publicly available at the data repository \textit{zenodo} {(\href{https://zenodo.org/doi/10.5281/zenodo.10696697}{10.5281/zenodo.10696697})}.

\begin{appendices}
\section{Free energy contributions}\label{sec:appendixA}
Here, we briefly discuss the individual contributions to the energy density, i.e., to the integrand of the {energy functional $\mathcal{F}$, i.e., Eq.~\eqref{eq:FullFreeEnergy} of the main text.} 
First, we consider the brush energy per substrate area $f_\mathrm{b}$. 
As the brush consists of end-tethered polymer chains without cross-coupling, we use an amended Flory-Huggins free energy of mixing. In particular, as compared to a standard model of a ternary mixture the translational entropy of the polymers is replaced by an entropic spring term ultimately employing an Alexander-de~Gennes~approach \cite{Alex1977jp,Genn1991crasi}, as discussed in Ref.~\cite{SmRB2020m,Somm2017m}. 
Writing the expression in terms of $\zeta$'s we have
\begin{align}\label{eq:fBrush}
	f\tb{b}= kT&\left[\rho_{\mathrm{l}1}\zeta_{\mathrm{b}1} \ln\frac{\zetB }{\zetB +\zetE +\zetP}
	+ \rho_{\mathrm{l}2}\zeta_{\mathrm{b}2} \ln\frac{\zetE }{\zetB +\zetE +\zetP } 
	+ \frac{\zetP \cdot(\chi_{1\mathrm{p}}\rho_{\mathrm{l}1}\zeta_{\mathrm{b}1} +\chi_{2\mathrm{p}}\rho_{\mathrm{l}2}\zeta_{\mathrm{b}2})}{\zetB +\zetE +\zetP } 
	\right.
	\nonumber\\
	&\,\,\left.
	+ \chi_{12}\frac{ \rho_{\mathrm{l}1}\zeta_{\mathrm{b}1}  \zetE}{\zetB +\zetE +\zetP } 
	+ \frac{3\sigma^2 }{2\zetP l_\mathrm{K}^3}(\zetB +\zetE +\zetP )^2\right] + c_{\mathrm{l}1} \zetB + c_{\mathrm{l}2} \zetE
\end{align}
where $\sigma$ is the relative grafting density while $\chi_{12}, \chi_{1\mathrm{p}},$ and $\chi_{2\mathrm{p}}$ are the Flory-Huggins $\chi$-parameters for the interactions of liquid~1 with liquid~2, liquid~1 with polymer, and liquid~2 with polymer, respectively. The final two terms account for the liquid bulk energy as compared to vapor bulk, and are relevant for evaporation/condensation (cf.~Ref.~\cite{HDJT2023jfm}), i.e., for the direct exchange between brush and gas. This ensures full consistency of the description of the drop energy with the limiting case of a fully saturated brush. Note that the terms are not present in Ref.~\cite{KHHB2023jcp}. The parameters $c_{\mathrm{l}1}$ and $c_{\mathrm{l}2}$ are constant energies per volume for the two pure liquids. They may be expressed in terms of relative humidities (see below).

The liquid bulk energy per substrate area $f\tb{d}$ is similar to the brush energy in the limit without polymers $\zetP, \sigma\to0$, i.e.,
\begin{align}\label{eq:fLiquid}
	f\tb{d}=kT\left[\rho_{\mathrm{l}1}\zeta_{\mathrm{d}1}\ln\frac{\zetA}{\zetA+\zetD} + \rho_{\mathrm{l}2}\zeta_{\mathrm{d}2}\ln\frac{\zetD}{\zetA+\zetD} + \chi_{12}\frac{\rho_{\mathrm{l}1}\zeta_{\mathrm{d}1}\zetD}{\zetA+\zetD}\right] + c_{\mathrm{l}1} \zetA + c_{\mathrm{l}2} \zetD\,,
\end{align}
i.e., it combines standard Flory-Huggins mixing energy with the two bulk liquid energies.

Within the gas phase we consider both components to be of sufficiently low density to be described as ideal gases. 
Following \cite{HDJT2023jfm} we assume the total gas pressure $\rho\tb{tot}$ to be constant and uniform on the time scale of the considered processes. Therefore, the number density (per volume) of the air the vapors diffuse in is
\begin{align}\label{eq:rhos}
	\rho\tb{air}(\vec{r},t) =\rho\tb{tot} - \rhoC - \rhoF = \rho\tb{tot} - \frac{\psi_3(\vec{r},t)}{h\tb{v}(\vec{r},t)} - \frac{\psi_6(\vec{r},t)}{h\tb{v}(\vec{r},t)}\,,
\end{align}
where $h\tb{v}=d-h\tb{d}-h\tb{b}$ is the local vertical extension of the gas phase (cf.~Fig.~\ref{fig:full-system-sketch}). 
Thus, the gas bulk energy is
\begin{align}
	f\tb{g} = kT \sum_{j=1}^2 \rho_{\text{v}j}\ln(\Lambda^3\rho_{\text{v}j}) + kT\rho\tb{air}\ln(\Lambda^3\rho\tb{air}) - kT\rho_\mathrm{tot}
\end{align}
where $\Lambda$ is the standard thermal de Broglie wavelength.
Considering a macroscopically thick flat layer of a pure liquid 1 or 2 in coexistence with saturated vapor, the influence of wetting and interface energies on the equilibrium state can be neglected. Then, the condition of equal chemical potentials reduces to 
\begin{align}
	\frac{\partial f\tb{d}}{\partial\psi_{i}} 		+
	\frac{\partial (h\tb{v}f\tb{g})}{\partial\psi_{i}}
	&= \frac{\partial (h\tb{v}f\tb{g})}{\partial \psi_{i+2}}  &&\text{for } i=1,4\,. \label{eq:dropVaporFlux}
\end{align}
As normally $\rho_\mathrm{tot}\ll \rho_{\mathrm{l}i}$ we may omit the second term on the left hand side allowing us to express the constants $c_{\mathrm{d}i}$ in Eq.~\eqref{eq:fLiquid} as $c_{\mathrm{d}i}=kT\rho_{\mathrm{l}i}\ln[\rho_{\mathrm{v}i}^\mathrm{sat}/(\rho\tb{tot}-\rho_{\mathrm{v}i}^\mathrm{sat})]$ as in section 2.2.4 of \cite{HDJT2023jfm}, i.e., in terms of number densities of saturated pure vapor $\rho_{\mathrm{v}i}^\mathrm{sat}$ which are parameters in our model. 

After introducing the bulk energies for the three phases we turn to the energies of the interfaces between them. Macroscopically, brush and/or mixture state-dependent interface energies are assigned to brush-drop ($\gamma_\mathrm{bd}$), drop-gas ($\gamma_\mathrm{dg}$), and brush-gas ($\gamma_\mathrm{bg}$) interfaces. However, as we employ a mesoscopic description of the system only energies $\gamma_\mathrm{bd}$ and $\gamma_\mathrm{dg}$ enter while $\gamma_\mathrm{bg}$ is encoded in the employed wetting energy $f_\mathrm{wet}$. As the gas is of low density 
$\gamma_\mathrm{dg}$ and $\gamma_\mathrm{bg}$ are independent of its state. For all other dependencies we assume a linear interpolation (as a function of liquid concentrations) between known limiting cases involving pure liquids. For instance, $\gamma\tb{dg}$ linearly interpolates between the liquid~1-gas value $\gamma\tb{l1g}$ and the liquid~2-gas value $\gamma\tb{l2g}$ as
\begin{align}\label{eq:gammaLiquidGas}
	\gamma\tb{dg}=\phi\tb{d1}\gamma\tb{l1g} + \phi\tb{d2}\gamma\tb{l2g}\,,
\end{align}
where $\phi\tb{d1}=\zeta_{\mathrm{d}1}/h_\mathrm{d}$ and $\phi\tb{d2}=1-\phi\tb{d1}$  denote the respective local vertically averaged volume fraction of liquid~1 and liquid~2 within the drop. 
For the brush-drop interface energy more limiting cases have to be accounted for, as summarized in Tab.~\ref{tab:limitingCasesLiqBrush}. 
\begin{table}[h]
	\centering
	\renewcommand{\arraystretch}{2.5}
	\caption{All considered limiting cases for the brush-drop interface energy $\gamma_\mathrm{bd}$.
		\label{tab:limitingCasesLiqBrush}}
	\begin{tabular}{|p{0.35\textwidth}|p{0.25\textwidth}|p{0.25\textwidth}|}
		\hline
		\diagbox{brush state}{drop state} & pure liquid~1 ($\phi\tb{d1}=1$) & pure liquid 2 ($\phi\tb{d2}=1$)\\\hline
		dry brush $(\phi_\mathrm{bp}=1)$ 						& $\gamma\tb{pl1}$					&	$\gamma\tb{pl2}$ \\
		saturated  with liquid~1 $(\phi\tb{b1}=1)$ 	& 0									& $\gamma_{12}\ll \gamma\tb{pl2}$ \\
		saturated  with liquid~2 $(\phi\tb{b2}=1)$ 	& $\gamma_{12}\ll\gamma\tb{pl1}$									& 0\\\hline
	\end{tabular}
\end{table}
As the two liquids are miscible or weakly immiscible (depending on $\chi_{12}$), the energy $\gamma_{12}$ is comparatively small. Linear interpolation gives
\begin{align}\label{eq:gammaLiquidBrush}
	\gamma\tb{bd}=\phi_\mathrm{bp}\phi\tb{d1} \gamma\tb{pl1} + \phi_\mathrm{bp}\phi\tb{d2} \gamma\tb{pl2} 
	+ \left[ \phi\tb{b1} \phi\tb{d2} + \phi\tb{b2} \phi\tb{d1}\right] \gamma_{12}\,.
\end{align}
Similarly, the macroscopic brush-gas interface energy is
\begin{align}\label{eq:gammaBrushGas}
	\gamma\tb{bg}=\phi\tb{b1}\gamma\tb{l1g} + \phi\tb{b2}\gamma\tb{l2g} + \phi\tb{bp}\gamma\tb{pg}\,.
\end{align}
In Eqs.~\eqref{eq:gammaLiquidBrush} and~\eqref{eq:gammaBrushGas}, $\phi\tb{b1}$, $\phi\tb{b2}$ and $\phi\tb{bp}=(1-\phi\tb{b1}-\phi\tb{b2})$ with $\phi_{\mathrm{b}i}=\zeta_{\mathrm{b}i}/h_\mathrm{b}$ denote the local vertically averaged volume fraction of  liquid~1, liquid~2 and polymer within the brush, respectively.

The macroscopic energy $\gamma\tb{bg}$ enters the mesoscopic model
only implicitly via the definition of the wetting energy $f_\mathrm{wet}$. Here, we employ a simple expression commonly used for partially wetting liquids. It consists of the combination of two power laws \cite{Thie2010jpcm}
\begin{align}
f_\mathrm{wet}(h\tb{d}) =  \frac{A}{2h\tb{d}^2}\, \left(\frac{2h\tb{a}^3}{5h\tb{d}^3}-1\right) 
\end{align}
where $h\tb{a}$ is an adsorption layer height that, for simplicity, is assumed constant. However, the Hamaker ``constant'' $A$ depends on the various concentrations and is determined via the consistency condition relating mesoscopic and macroscopic description \cite{TSTJ2018l, HDGT2024l}, namely, $\gamma\tb{bg} = \gamma\tb{bd} + \gamma\tb{dg} + f_\mathrm{wet}(h\tb{a})$. As $f_\mathrm{wet}(h\tb{a})=-3A/10h\tb{a}^2$, this implies
\begin{align}
A = -\frac{10}{3}h\tb{a}^2 \,\left(\gamma\tb{bg} - \gamma\tb{bd} - \gamma\tb{dg}\right)
\end{align}
where the $\gamma$'s are given by \eqref{eq:gammaLiquidBrush}, \eqref{eq:gammaLiquidGas}, and \eqref{eq:gammaBrushGas}. 

Having defined all terms contributing to the energy functional \eqref{eq:FullFreeEnergy} we emphasize that the dependencies written as functions of effective thicknesses (the $\zeta$'s) and volume fractions (the $\phi$'s) have to be expressed in terms of particle number densities (the $\psi$'s) before performing the variations entering the gradient dynamics model \eqref{eq:gradDyn}. This is cumbersome but straightforward.

\section{Mobilities}\label{sec:appendixB}
Finally, we briefly discuss the $6\times6$ mobility matrices $\tens{Q}$ and $\tens{M}$ that encode the various dissipation channels associated with the conserved and nonconserved dynamics, respectively, in the gradient dynamics model \eqref{eq:gradDyn}. As we base the sequence of fields in the state vector $(\psi_1,\psi_2,\psi_3,\psi_4,\psi_5,\psi_6)$ first on substance and second on phase (drop, brush and gas), both matrices can be written in block form using $3\times3$ sub-matrices, namely, $\tens{Q}=((\tens{Q}_1, \tens{Q}_\mathrm{mix}),(\tens{Q}^\mathrm{T}_\mathrm{mix}, \tens{Q}_2))$ and $\tens{M}=((\tens{M}_1, \tens{M}_\mathrm{mix}),(\tens{M}^\mathrm{T}_\mathrm{mix}, \tens{M}_2))$. The particular expressions are all based on known limiting cases. We emphasize that positive definiteness and symmetry of the matrices ensure thermodynamic consistency (corresponding to Onsager reciprocal relations between transport coefficients).

First, we consider the conserved part. Within the drop we use the established mobilities for a film of binary liquid mixture written in symmetric form~\cite{XuTQ2015jpcm}, while in the brush and the gas phase we consider only diffusive transport. In the gas, as expected one obtains the standard diffusion equation in the limit $\rho\tb{air}\gg\rhoC,\rhoF$. However, in the brush the assumption of transport by diffusion only is a rather strong approximation strictly valid for large $\phi_{\mathrm{bp}}$ only. Here, we use it as a sensible first approximation that gives rather reasonable results as shown for simple liquids \cite{HDGT2024l}. Considering within brush and gas only by diffusion implies that there is no dynamic coupling between the three phases. Thus, we obtain 
\begin{align}
	\tens{Q}_i = 
	\begin{pmatrix}
		\frac{(\zetA+\zetD)\psiDi^2}{3\eta} + \rho^2_{\mathrm{l}i}\frac{\zetA\zetD}{\zetA+\zetD}\frac{D_\mathrm{d}}{kT} & 0 &0\\
		0 & \rho^2_{\mathrm{l}i}\zeta_{\mathrm{b}i}\frac{D_\mathrm{b}}{kT} & 0 \\
		0 & 0 & \psi_{\mathrm{v}i}\frac{D_\mathrm{v}}{kT}
	\end{pmatrix} 
\quad \mathrm{ for }\: i=1,2\,,
\end{align}
for the block matrices on the diagonal of $\tens{Q}$, and 
\begin{align}
	\tens{Q}_\mathrm{mix} = 
	\begin{pmatrix}
		\frac{(\zetA+\zetD)\psiA\psiD}{3\eta} - \rho_{\mathrm{l}1}\rho_{\mathrm{l}2}\frac{\zetA\zetD}{\zetA+\zetD}\frac{D_\mathrm{d}}{kT} & 0 &0\\
		0 & 0 & 0 \\
		0 & 0 & 0
	\end{pmatrix}
\end{align}
for the off-diagonal block matrix. Here, $D_\mathrm{d}$, $D_\mathrm{b}$ and $D_\mathrm{v}$ denote the diffusion coefficients in the drop, brush and gas phase, respectively. Note that cross-diffusion is neglected in brush and gas.

Considering the nonconserved dynamics, the matrix $\tens{M}$ has no off-diagonal part ($\tens{M}_\mathrm{mix}=\tens{0}$) because substances 1 and 2 do not chemically transform into each other. The block matrices on the diagonal are 
\begin{align}
	\tens{M}_i = 
	\begin{pmatrix} 
		{ M_i^\mathrm{db}+M_i^\mathrm{dg}} &-M_i^\mathrm{db}  &-M_i^\mathrm{dg}  
		\\
		-M_i^\mathrm{db}  & M_i^\mathrm{db} + {M}_i^\mathrm{bg}&-{M}_i^\mathrm{bg}
		\\
		-M_i^\mathrm{dg}& -{M}_i^\mathrm{bg}	&M_i^\mathrm{dg} + {M}_i^\mathrm{bg}
	\end{pmatrix}
	\quad \mathrm{ for }\: i=1,2\,.
\end{align}
They are identical to the corresponding matrix for a one-component brush-drop-gas system discussed in Ref.~\cite{HDGT2024l}. In particular, $M_i^\mathrm{dg}$ and ${M}_i^\mathrm{bg}$ represent the evaporation rates of liquid from drop to gas and brush to gas, respectively (negative rates indicate condensation). Furthermore, $M_i^\mathrm{db}$ is the imbibition rate from drop into brush (negative rate indicates desiccation). Note that the sums of all components add up to zero for each row and column. This indicates material conservation across the overall system, i.e., summing up the respective first and second three equations of the model \eqref{eq:gradDyn} results in two conservation laws.

\section{Parameters}\label{sec:appendixC}
Here, we briefly present the parameters used in {section}~\ref{sec:results}.  An effective nondimensionalization of the dynamic equations is obtained by setting $\eta,\,kT,\,h_\mathrm{a},\,\rho_{\mathrm{l}1}=1$ and using an according rescaling of all other parameters. Therefore all parameter values given here and in the main text are nondimensional. 
In particular, the brush-specific parameters are the grafting density $(\sigma, N)=(0.2,75)$, the Flory-Huggins $\chi$-parameters $(\chi_\mathrm{12}, \chi_\mathrm{1p}, \chi_\mathrm{2p})=(0.1,0.25,0.85)$. The reference interface energies are $(\gamma_\mathrm{pl1}, \gamma_\mathrm{pl2},\gamma_\mathrm{pg},\gamma_\mathrm{12}, \gamma_\mathrm{l1g}, \gamma_\mathrm{l2g})=(14,15,20,0.05,5,6)$. In the conserved and nonconserved mobilities we have $(D\tb{d}, D\tb{b}, D\tb{v})=(0.005,0.001,0.1)$ and $(M_1^\mathrm{db}=M_1^\mathrm{dg}=M_1^\mathrm{bg}, M_2^\mathrm{db}=M_2^\mathrm{dg}=M_2^\mathrm{bg})=(1\cdot10^{-4},2\cdot10^{-4})$, respectively.

Furthermore, for simplicity we use $l\tb{K}=\Lambda=h_\mathrm{a}$ and $10^{3}\rho_{\mathrm{tot}}=10^{5}\rho^\mathrm{sat}_{\mathrm{v}1}=10^{5}\rho^\mathrm{sat}_{\mathrm{v}2}=\rho_{\mathrm{l}2}=\rho_{\mathrm{l}1}$. The identical vapor saturation densities $\rho^\mathrm{sat}_{\mathrm{v1}}=\rho^\mathrm{sat}_{\mathrm{v2}}$ for the two pure liquids are denoted by $\rho^\mathrm{sat}_{\mathrm{v}}$. In consequence, $c_{\mathrm{d}1}=c_{\mathrm{d}2}=-4.6$.

After reformulation in terms of densities $\rho=p/kT$ and only considering our choice $\rho_{\mathrm{l}2}=\rho_{\mathrm{l}1}$, Raoult's law \cite{AtkinsPaula2010} becomes $ (\rhoC+\rhoF)/\rho^\mathrm{sat}_{\mathrm{v}} = \phi_{\mathrm{d}1} + \phi_{\mathrm{d}2}=1$. As Raoult's law only holds for ideal mixtures we see a small deviation as mentioned in {section}~\ref{sec:results}. 

\end{appendices}

\end{document}